\documentclass[proceedings]{JHEP} 
\usepackage{epsfig}			
\def\slashchar#1{\setbox0=\hbox{$#1$}           
   \dimen0=\wd0                                 
   \setbox1=\hbox{/} \dimen1=\wd1               
   \ifdim\dimen0>\dimen1                        
      \rlap{\hbox to \dimen0{\hfil/\hfil}}      
      #1                                        
   \else                                        
      \rlap{\hbox to \dimen1{\hfil$#1$\hfil}}   
      /                                         
   \fi}                                         %

\title{Decays and Lifetime of $\bf B_c$ in QCD Sum Rules}
\author{V.V.Kiselev, A.E.Kovalsky, A.K.Likhoded\\
        Russian State Research Center "Institute for High Energy
        Physics",\\ Protvino, Moscow Region, 142284 Russia\\
		Email: \email{kiselev@th1.ihep.su}}
\conference{Heavy Quark Physics 5, Dubna, Russia, 6-8 April 2000}
\abstract{In the framework of three-point QCD sum rules, the form factors for
the semileptonic decays of $B_c^+\rightarrow B_s(B_s^*) l^+\nu_l$ are
calculated with account for the Coulomb-like $\alpha_s/v$-corrections in the
heavy quarkonium. The generalized relations due to the spin symmetry of
HQET/NRQCD for the form factors are derived at the recoil momentum close to
zero. The nonleptonic decays are studied using the assumption on the
factorization. The $B_c$ meson lifetime is estimated by summing up the
dominating exclusive modes in the $c \rightarrow s$ transition combining the
current calculations with the previous analysis of $b \rightarrow c$ decays in
the sum rules of QCD and NRQCD.}
\begin{document} 

\section{Introduction}
For better understanding and precise measuring the weak-action properties of
heavy quarks, governed by the QCD forces, we need as wide as possible
collection of snapshots with hadrons, containing the heavy quarks. Then we can
provide the study of heavy quarks dynamics by testing the various conditions,
determining the forming of bound states as well as the entering of strong
interactions into the weak processes. So, a new lab for such investigations is
a doubly heavy long-lived quarkonium $B_c$ recently observed by the CDF
Collaboration \cite{cdf} for the first time. 

This meson is similar to the charmonium and bottomonium in the spectroscopy,
since it is composed by two nonrelativistic heavy quarks, so that the NRQCD
approach \cite{NRQCD} is well justified to the system. The modern predictions
for the mass spectra of $\bar b c$ levels were obtained in refs. \cite{eichten}
in the framework of potential models and lattice simulations. The measured
value of $B_c$ mass yet has a large uncertainty
$
M_{B_c} = 6.40\pm 0.39\pm 0.13\; {\rm GeV,}
$
in agreement with the theoretical expectations.

The measured $B_c$ lifetime
$$
\tau[B_c] = 0.46^{+0.18}_{-0.16}\pm 0.03\; {\rm ps,}
$$
agrees with the estimates obtained in the framework of both the OPE combined
with the evaluation of hadronic matrix elements in NRQCD
\cite{Bigi,Buchalla,BcOnish}
and potential quark models, where one has to sum  up the dominating exclusive
modes to calculate the total $B_c$ width \cite{LusMas,Bcstatus},
$
\tau_{\rm OPE,PM}[B_c] = 0.55\pm 0.15\; {\rm ps.}
$
The accurate measurement of $B_c$ lifetime could allow one to distinguish
various parameter dependencies such as the optimal heavy quark masses, which
basically determine the theoretical uncertainties in OPE. 

At present, the calculations of $B_c$ decays in the framework of QCD sum rules
were performed in \cite{Colangelo,Bagan,KT,Onish}. The authors of
\cite{Colangelo,Bagan} got the results, where the form factors are about 3
times less than the values expected in the potential quark models, and the
semileptonic and hadronic widths of $B_c$ are one order of magnitude less than
those in OPE. The reason for such the disagreement was pointed out in \cite{KT}
and studied in \cite{Onish}: in the QCD sum rules for the heavy quarkonia the
Coulomb-like corrections are significant, since they correspond to summing up
the ladder diagrams, where $\alpha_s/v$ is not a small parameter, as the heavy
quarks move nonrelativistically, $v\ll 1$. The Coulomb rescaling of
quark-quarkonium vertex enhances the estimates of form factors in the QCD sum
rules for the $B_c^+\to \psi(\eta_c) l^+ \nu$ decays. In \cite{Onish} the soft
limit $v_1\cdot v_2\to 1$ at $v_1\neq v_2$, where $v_{1,2}$ denote the
four-velocities of initial and recoil mesons, was considered, and the
generalized spin symmetry relations were obtained for the $B_c\to\psi(\eta_c)$
transitions: four equations, including that of \cite{Jenkins}. Moreover, the
gluon condensate term was calculated in both QCD and NRQCD, so that it enforced
a convergency of the method.

In the present paper we calculate the $B_c$ decays due to the $c\to s$ weak
transition in the framework of QCD sum rules, taking into account the
Coulomb-like $\alpha_s/v$-corrections for the heavy quarkonium in the initial
state. In the semileptonic decays the hadronic final state is saturated by the
pseudoscalar $B_s$ and vector $B_s^*$ mesons, so that we need the values of
their leptonic constants entering the sum rules and determining the
normalization of form factors. For this purpose, we reanalyze the two-point sum
rules for the $B$ mesons to take into account the product of quark and gluon
condensates in addition to the previous consideration of terms with the quark
and mixed condensates. We demonstrate the significant role of the product term
for the convergency of method and reevaluate the constants $f_B$ as well as
$f_{B_s}$. Taking into account the dependence on the threshold energy $E_c$ of
hadronic continuum in the $\bar b s$ system in both the value of $f_{B_s}$
extracted from the two-point sum rules and the form factors in the three-point
sum rules, we observe the stability of form factors versus $E_c$, which
indicates the convergency of sum rules.

The spin symmetries of leading terms in the lagrangians of HQET \cite{HQET} for
the singly heavy hadrons (here $B_s^{(*)}$) and NRQCD \cite{NRQCD} for the
doubly heavy mesons (here $B_c$) result in the relations between the form
factors of semileptonic $B_c\to B_s^{(*)}$ decays. We derive two generalized
relations in the soft limit $v_1\cdot v_2\to 1$: one equation in addition to
what was found previously in ref.\cite{Jenkins}. The relations are in a good
agreement with the sum rules calculations up to the accuracy better than 10\%,
that shows a low contribution of next-to-leading $1/m_Q$-terms.

We perform the numerical estimates of semi\-leptonic $B_c$ widths and use the
factorization approach \cite{blokshif} to evaluate the nonleptonic modes.
Summing up the dominating exclusive modes, we calculate the lifetime of $B_c$,
which agree with the experimental data and the predictions of OPE and quark
models. We discuss the preferable prescription for the normalization point of
nonleptonic weak lagrangian for the charmed quark and present our optimal
estimate of total $B_c$ width. We stress that in the QCD sum rules to the given
order in $\alpha_s$, the uncertainty in the values of heavy quark masses is
much less than in OPE. This fact leads to a more definite prediction on the
$B_c$ lifetime.

\section{Three-point sum rules}

The hadronic matrix elements for the semileptonic $B_c(p_1)\to B_s(p_2)$ decays
can be written down as follows: 
\begin{eqnarray}
&&\langle B_s|V_{\mu}|B_c\rangle = f_{+}(p_1 + p_2)_{\mu} +
f_{-}q_{\mu},\\
&&\frac{1}{i}\langle B_s^* |V_{\mu}|B_c\rangle = 
i F_V\epsilon_{\mu\nu\alpha\beta}\epsilon^{*\nu}(p_1 +
p_2)^{\alpha}q^{\beta},\nonumber\\
&&\frac{1}{i}\langle B_s^* |A_{\mu}|B_c\rangle =
F_0^A\epsilon_{\mu}^{*} + 
F_{+}^{A}(\epsilon^{*}\cdot p_1)(p_1 + p_2)_{\mu} \nonumber \\ &&
\hspace*{26mm}+ F_{-}^{A}(\epsilon^{*}\cdot p_1)q_{\mu}, \nonumber
\end{eqnarray}
where $q_{\mu} = (p_1 - p_2)_{\mu}$ and $\epsilon^{\mu} = \epsilon^{\mu}(p_2)$
is the polarization vector of $B_s^*$ meson. $V_{\mu}$ and $A_{\mu}$ are the
flavour changing vector and axial electroweak currents. Following
the standard procedure for the evaluation of form factors in the framework of
QCD sum rules \cite{SR3pt}, we consider the three-point functions, say,
\begin{eqnarray}
&&\Pi_{\mu}(p_1, p_2, q^2) = i^2 \int dxdye^{i(p_2\cdot x - p_1\cdot
y)} \cdot \nonumber\\
&&\langle 0|T\{\bar q_1(x)\gamma_5 q_2(x), V_{\mu}(0), \bar b(y)\gamma_5
c(y)\}|0\rangle,\nonumber
\end{eqnarray}
where $\bar q_1(x)\gamma_5 q_2(x)$ and $\bar q_1(x)\gamma_{\nu}q_2(x)$ denote
interpolating currents for $B_{s}$ and $B_{s}^*$, correspondingly. 

The Lorentz structures in the correlators can be written down as
$
\Pi_{\mu} = \Pi_{+}(p_1 + p_2)_{\mu} + \Pi_{-}q_{\mu}.
$
The form factors $f_{\pm}$ are determined
from the amplitudes $\Pi_{\pm}$, respectively. 

The leading QCD term is a triangle quark-loop diagram,
for which we can write down the double dispersion representation at
$q^2\leq 0$
\begin{eqnarray}
&&\Pi_i^{pert}(p_1^2, p_2^2, q^2) = -\frac{1}{(2\pi)^2}\cdot \nonumber \\
&&\int \frac{\rho_i^{pert}(s_1, s_2, q^2)}{(s_1 - p_1^2)(s_2 - p_2^2)}ds_1ds_2
+ \mbox{subtractions}, \nonumber
\end{eqnarray}
where the limits of integration region and the spectral densities
are given in \cite{Onish}.

The physical spectral functions are generally saturated by the ground hadronic
states and a continuum starting at some effective thresholds.

\setlength{\unitlength}{0.5mm}
\vspace*{2mm}
\begin{figure}[th]
\begin{center}
\begin{picture}(110, 100)
\put(0, 0){\epsfxsize=5.7cm \epsfbox{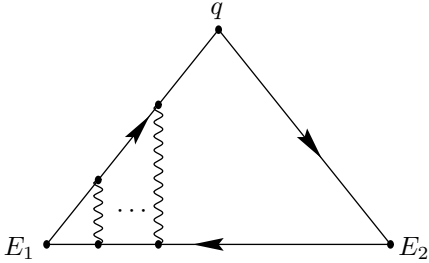}}
\put(0,40){$E_1$}
\put(105,40){$E_2$}
\put(55,104){$q$}
\put(30,50){$\cdots$}
\end{picture}
\end{center}
\normalsize

\vspace*{-2cm}
\caption{The ladder diagram of the Coulomb-like interaction.} 
\label{Coul-fig}
\end{figure}

For the heavy quarkonium $\bar b c$, where the relative velocity of quark
movement is small, an essential role is taken by the Coulomb-like
$\alpha_s/v$-corrections. They are caused by the ladder diagram, shown in
Fig. \ref{Coul-fig}. This leads to the finite renormalization for $\rho_i$
\cite{Onish}, so that
$
 \rho^{c}_i={\bf C} \rho_i,
$
$$
 {\bf C}=\frac{|\Psi^C_{\bar b c}(0)|}{|\Psi^{free}_{\bar b
 c}(0)|}=\sqrt{\frac{4\pi
 \alpha_s}{3v}(1-\exp\{-\frac{4\pi \alpha_s}{3v}\})^{-1}},
$$
where $v$ is the relative velocity of quarks in the $\bar b c$-system,
$
v=\sqrt{1-\frac{4 m_b m_c}{p_1^2-(m_b-m_c)^2}}.
$

\section{Numerical estimates}
We evaluate the form factors in the scheme of spectral density moments. This
scheme is not strongly sensitive to the value of the $\bar b c$-system
threshold energy, and we put $E_c^{\bar b c}=1.2~\mbox{GeV}$. The
two-point sum rules for the $B_c$ meson with account for the Coulomb-like
corrections give $$\alpha_s^{c}(\bar b c)=0.45,$$ which corresponds to
$f_{B_c}$=400 MeV \cite{fbc}. The quark masses are fixed by the calculations of
leptonic constants $f_{\Psi}$ and $f_{\Upsilon}$ in the same order over
$\alpha_s$. The requirement of stability in the sum rules  including the
contributions of higher excitations, results in quite an accurate determination
of masses $m_c=1.40\pm0.03~$GeV and $m_b=4.60\pm0.02~$GeV, which are in a good
agreement with the recent estimates in \cite{masses}, where the quark masses
free off a renormalon ambiguity were introduced. 

The leptonic constant for the $B_s$ meson is extracted from the two-point sum
rules. The Borel improved sum rules for the $B$ meson leptonic constant
\cite{Neubert} have the following form:
\begin{eqnarray}
&& f^2_B M_Be^{-\bar \Lambda(\mu) \tau}=K^2 \frac{3}{\pi^2} C(\mu)
 \int\limits^{\omega_0(\mu)}_{0}
 d\omega~\omega^2 e^{-\omega \tau}\nonumber \\ &&
 +\langle \bar q q \rangle(1-
 \frac{m_0^2~\tau^2}{16}+ \frac{\pi^2 \tau^4}{288}\langle\frac{\alpha_s}{\pi}
 G^2 \rangle), \nonumber
\end{eqnarray}
where the K-factor is due to $\alpha_s$-corrections \cite{Neubert}.
We find that NLO corrections to the leptonic constant are about $40\%$. Using
the Pad\'e approximation, we find that higher orders corrections can be about
$30\%$. So, we hold the K factor in conservative limits $1.4\div1.7$. It is
quite reasonable to suppose its cancellation in evaluating the semileptonic
form factors due to the renormalization of heavy-light vertex in the triangle
diagram.
In the limit of semi-local duality \cite{fesr,SU(3)} $\tau \rightarrow 0$ we
get the relation: $\bar \Lambda(\mu) = \frac{3}{4}~\omega_0(\mu)$. We introduce
the renormalization invariant quantities
$
\omega_{0,dual}^{ren}=C^{-1/3}(\mu)~\omega_0(\mu),\;
\bar\Lambda^{ren}_{dual} = \frac{3}{4}~\omega_{0,dual}^{ren}.
$ 
For $\bar\Lambda^{ren}_{dual}$ we have $\bar \Lambda^{ren}_{dual} = M_B-m_b =
0.63~\mbox{GeV}$, and we obtain that in the semi-local duality the threshold
energy $~\omega_{0,dual}^{ren}=0.84~\mbox{GeV}$. Neglecting the quark
condensate term in the leptonic constant we have 
$ 
f_B^2M_B=K^2 \frac{3}{\pi^2}(\omega_{0,dual}^{ren})^3.
$
In the general Borel scheme for $f_B$ we have to consider the stability at
$\tau \neq0$ with the extended region of resonance contribution. We expect,
that the sum rules  with the redefined $\omega^{ren}$ and $\bar \Lambda^{ren}$
have a stability point at $\tau \sim \frac{1}{\bar \Lambda}$. The results are
in a good agreement with the semi-local duality if the threshold energy of
continuum equals $E_c=1.1\div1.3~$GeV (see Fig. \ref{stabil}, where the overall
K-factor was ignored). So, we find the $E_c^{3/2}$-dependence of $f_B
\sqrt{M_B}$, whereas the contribution of condensate is numerically suppressed,
as expected from the semi-local duality.
Multiplying the result taken from Fig. \ref{stabil}, by the
K-factor we find the value $f_B=140\div170~$MeV, which is in a good agreement
with the recent lattice results \cite{Lellouch} and the estimates in the QCD SR
by other authors \cite{Braun}. 

\FIGURE{$\hspace*{-30mm}f_B\sqrt{M},\; {\rm
GeV}^{3/2}\hspace*{5cm}\bar\Lambda,\; {\rm
GeV}$\\
\hspace*{-11mm}\epsfig{file=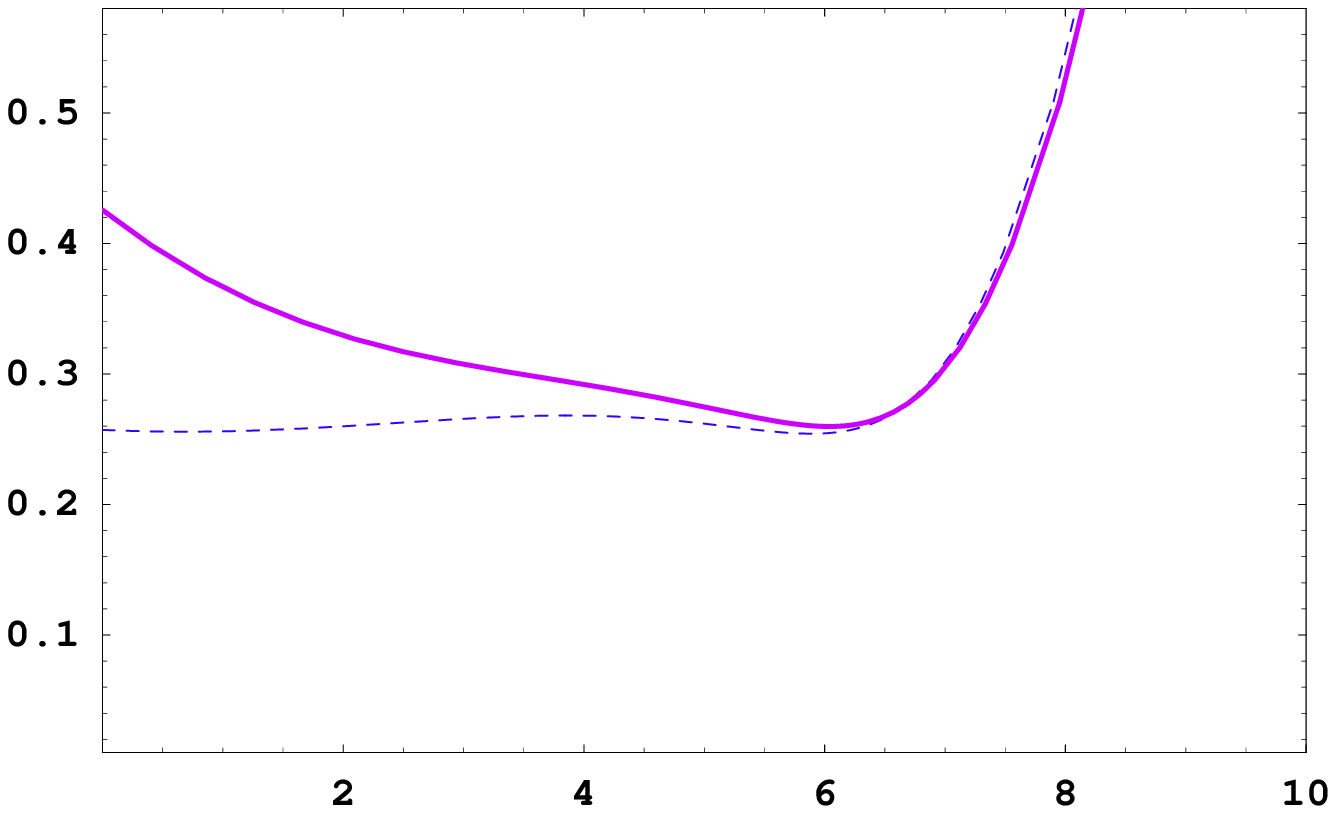,width=6.7cm}
\epsfig{file=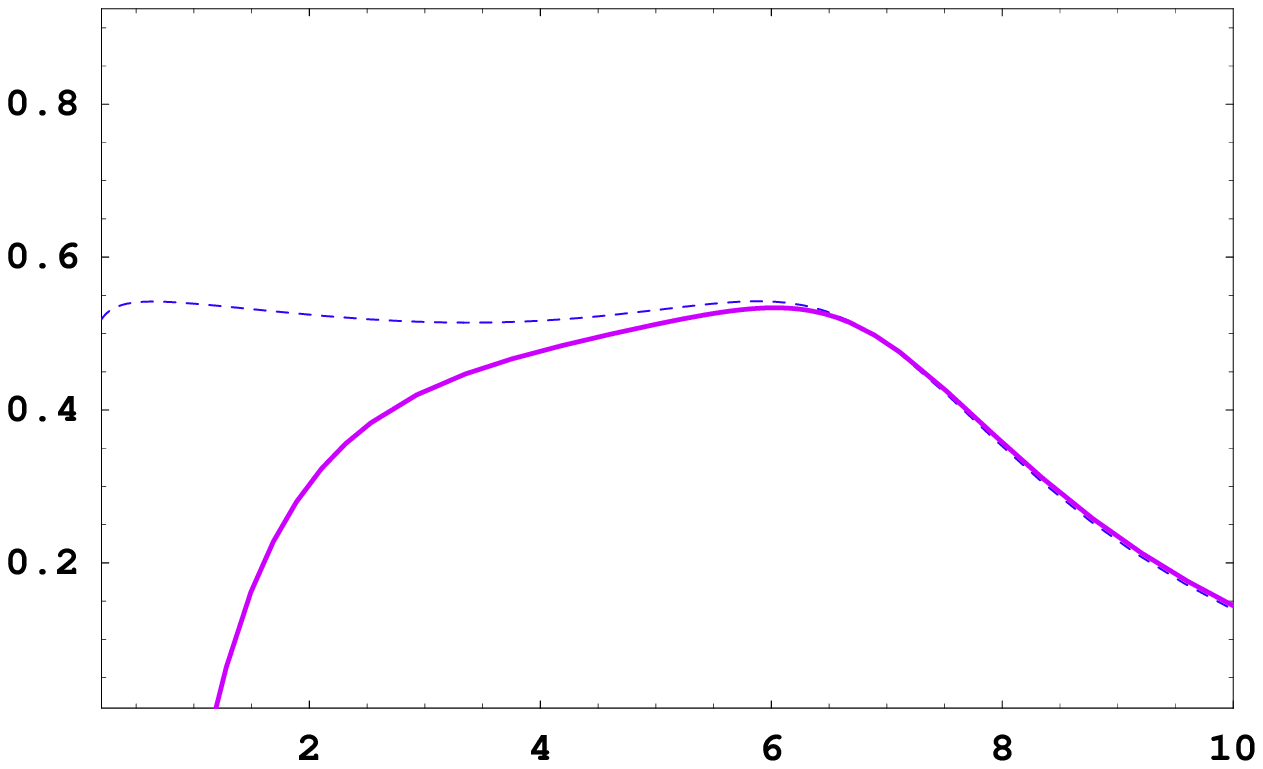,width=6.7cm}\\ 
$\hspace*{50mm}\tau,\; {\rm GeV}^{-1}\hspace*{6cm}\tau,\; {\rm GeV}^{-1}$
        \caption{The leptonic constant of B meson and the b-quark binding
        energy in the semi-local duality sum rules (dashed curve)
        and in the general Borel scheme (solid line) with the corrected value
        of $\bar \Lambda$, which improves the stability of result obtained in
        the semi-local duality.} 
\label{stabil}}

For the vector $B^*$ meson constant
$f_{B^*}$ we put $\frac{f_{B^*}}{f_{B}}=1.11~$(see \cite{Neubert2,Braun}). 
For the leptonic constant of $B_s$ meson we explore the following relation
$\frac{f_{B_s}}{f_B}=1.16$, which expresses the flavor SU(3)-symmetry violation
for B mesons \cite{SU(3)}.

We have investigated the dependence of form factors on the $\bar b s$ threshold
energy of continuum in the range $E_c=1.1 \div 1.3$ GeV, so that the optimal
choice for the $\bar b s$ system threshold energy is 1.2 GeV. In Table
\ref{form} we present the results of sum rules for the form factors. Comparing
with the estimates in the framework of potential models \cite{Bcstatus,Kis}, we
find a good agreement of estimates in the QCD sum rules with the values in the
quark model. 

\small
\TABLE{
\begin{tabular}{|c|c|c|c|}
\hline
$f_+$ & $f_-$ & $F_{V}, \mbox{GeV}^{-1}$ & $F_{0}^{A}, \mbox{GeV}$ 
 \\
\hline 
1.3&-5.8&1.1&8.1\\
\hline
\end{tabular}
\caption{The form factors of $B_{c}$ decay modes into the $B_{s}$ and
$B_{s}^{*}$ mesons at $q^{2}=0$.}
\label{form}
}
\normalsize 
The accuracy of sum rules under consideration is basically determined by the
variation of heavy quark masses. Indeed, the significant $\alpha_s$ correction
to the leptonic constant of $B_s$ meson should cancel the same factor for the
renormalization of quark-meson vertex in the triangle diagram. The dependence
on the choice of threshold energy in the $\bar b s$-channel can be optimized
and, hence, minimized. The variation of threshold energy in the $\bar b
c$-channel give the error less than 1\%. The effective coulomb constant is
fixed from the two-point sum rules for the heavy quarkonium, and its variation
is less than 2\%, which gives the same uncertainty for the form factors. The
heavy quark masses are determined by the two-point sum rules for the heavy
quarkonia, too. However, their variations result in the most essential
uncertainty. Summing up all of mentioned variations we estimate $\delta f/f
\simeq 5$\%.

The semileptonic widths calculated in the QCD sum rules are presented in Table
\ref{slwidth}. 

\TABLE{
\begin{tabular}{|c|c|c|}
\hline
 mode & $\Gamma$, $ 10^{-14}\ \mbox{GeV}$ & BR, $\%$ \\
\hline
 $B_{s}e^{+}\nu_{e}$ & 5.8 & 4.0 \\
\hline
 $B_{s}^{*}e^{+}\nu_{e}$ & 7.2 & 5.0 \\
\hline 
\end{tabular}
\caption{The widths of semileptonic $B_{c}$ decay modes and the
branching fractions calculated at $\tau_{B_{c}}=0.46$ ps.}
\label{slwidth}
}
\section{The symmetry relations}
At the recoil momentum close to zero, the heavy quarks in both the initial and
final states have small relative velocities inside the hadrons, so that the
dynamics of heavy quarks is essentially nonrelativistic. This allows us to use
the combined NRQCD/HQET approximation in the stu\-dy of mesonic form factors.
The expansion in the small relative velocities leads to
various relations between the form factors due to the spin symmetry
of effective lagragians to the leading order. Solving these relations
results in the introduction of an universal form factor (an analogue of the
Isgur-Wise function) at $q^2 \rightarrow q^2_{max}$.

We have derived the symmetry relations for the following form
factors:
\begin{eqnarray}
f_{+}(c_1^{P}\cdot{\cal M}_2 - c_2^{P}{\cal M}_1) - \hspace*{12mm}
&~&\nonumber \\
f_{-}(c_1^{P}\cdot{\cal
M}_2 + c_2^{P}\cdot{\cal M}_1) &=& 0,\nonumber\\
F_{0}^{A}\cdot c_V -2 c_{\epsilon}\cdot F_V{\cal M}_1{\cal M}_2 &=& 0,
\label{Fsym}\\
F_{0}^{A}c_1^{P} + c_{\epsilon}\cdot{\cal M}_1(f_{+} + f_{-}) &=& 0, \nonumber
\end{eqnarray}
where ${\cal M}_1=m_c+m_b$, ${\cal M}_2=m_s+m_b$, and
\begin{eqnarray}
c_{\epsilon} &=& -2,\; \nonumber\\
c_V &=& -1-\tilde B-\frac{m_b}{2m_c},\; \nonumber\\
c_1^{P} &=& 1-\tilde B+\frac{m_b}{2m_c},\; \\
c_2^{P} &=& 1+\tilde B-\frac{m_b}{2m_c}. \nonumber
\end{eqnarray}
Equating the second relation
in (\ref{Fsym}), for example, we obtain
$$
\tilde B=-\frac{2m_c+m_b}{2m_c}+\frac{4m_b(m_c+m_b)F_V}{F_0^A}\simeq 10.0,
$$
where all form factors are taken at $q^2_{max}$. Substituting $\tilde B$ in
first and third relations, we get $f_+\simeq 2.0$ and $f_-\simeq -8.3$. These
values have to be compared with the corresponding form factors obtained in the
QCD sum rules: $f_+(q^2_{max})=1.8$ and $f_-(q^2_{max})=-8.1$, where we suppose
the pole like behaviour of form factors. Thus, we find that in the QCD sum
rules, relations (\ref{Fsym}) are valid with the accuracy better than $10\%$ at
$q^2=q^2_{max}$. The deviation could increase at $q^2<q^2_{max}$ because of
variations in the pole masses governing the evolution of form factors. However,
in $B_c^+ \rightarrow B_s^{(*)}l^+\nu$ decays the phase space is restricted, so
that the changes of form factors are about 50\%, while their ratios develop
more slowly. 

\section{Nonleptonic decays and the lifetime}

The hadronic decay widths can be obtained on the basis of assumption on the
factorization for the weak transition between the quarkonia and the final
two-body hadronic states. For the nonleptonic decay modes the effective
Hamiltonian can be written down as 
$$
H_{eff}=\frac{G_F}{2 \sqrt{2}}V_{cs}V_{ud}^*\{C_+(\mu)O_{+}+C_-(\mu)O_-\},
\label{Heff}
$$
where 
$
O_{\pm}=(\bar u_i\gamma_{\nu}(1-\gamma_5)d_i)(\bar
s_j\gamma^{\nu}(1-\gamma_5)c_j) \pm (\bar u_i\gamma_{\nu}(1-\gamma_5)d_j)(\bar
s_i\gamma^{\nu}(1-\gamma_5)c_j),
$
and the factors $C_{\pm}(\mu)$ account for the strong corrections to the
corresponding four-fermion operators caused by hard gluons. The review on the
evaluation of $C_{\pm}(\mu)$ can be found in \cite{NLO}. The results are
collected in Table \ref{width}. 

\TABLE{
\begin{tabular}{|c|c|c|}
\hline
 mode &  $\Gamma$, $ 10^{-14}\ \mbox{GeV}$ & BR, $\%$ \\
\hline
 $B_{s}\pi^{+}$ & 15.8 $a_{1}^{2}$ & 17.5 \\
\hline
$B_{s}\rho^{+}$ & 6.7 $a_{1}^{2}$ & 7.4 \\
\hline
 $B_{s}^{*}\pi^{+}$ & 6.2 $a_{1}^{2}$ & 6.9 \\
\hline
 $B_{s}^{*}\rho^{+}$ & 20.0 $ a_{1}^{2}$ & 22.2 \\
\hline 
\end{tabular}
\caption{The widths of dominant nonleptonic $B_{c}$ decay modes due to $c
\rightarrow s$ transition and the
branching fractions calculated at $\tau_{B_{c}}=0.46$ ps. We put
$a_{1}$=1.26.}
\label{width}
}
In the parton approximation we could expect 
$ 
\Gamma[B_c^+ \rightarrow B_s^{(*)}+light~hadrons] = (2C_+^2(\mu)+C_-^2(\mu))
\Gamma[B_c^+ \rightarrow B_s^{(*)}e^+\nu_e], 
$
which results in the estimate very close to the value obtained as the sum of
exclusive modes at $\mu>0.9~$GeV. The deviation between these two estimates
slightly increase at $\frac{m_c}{2}<\mu<0.9~$GeV. Concerning the comparison of
hadronic width summing up the exclusive decay modes with the estimate based on
the quark-hadron duality, we insist that the deviation between these two
estimates is unessential since it is less that 10\%.

We estimate the lifetime using the fact that the dominant modes of the
$B_c~$meson decays are the $c \rightarrow s,~b \rightarrow c~$ transitions with
the $B_s^{(*)}$ and J/$\psi$, $\eta_c$ final states respectively, and the
electroweak annihilation \footnote{The $\bar b \rightarrow \bar c c \bar s$
transition is negligibly small in the $B_c$ decays because of destructive Pauli
interference for the charmed quark in the initial state and the product of
decay \cite{Buchalla}.}.

The method for the calculation of multi-par\-tic\-le branching fractions was
offered by Bjorken in his pioneering paper devoted to the decays of hadrons
containing heavy quarks \cite{Bjorken}. In order to estimate the contribution
of non-resonant $3\pi$ modes of $B_c$ decays into $B_s^{(*)}$ we use this
technique, i.e. the Poisson distribution with the average value
corrected to agree with the non-resonant $3\pi$-modes in the decays of $D$
mesons. We have found  ${\rm
BR}(B_c^+\to B_s^{(*)}(3\pi)^+)\approx 0.2$\%, while ${\rm BR}(B_c^+\to
B_s^{(*)}(2\pi)^+|_{\rm non-resonant})\approx 3$\%. We see that the neglected
modes contribute to the total width of $B_c$ as a small fraction in the limits
of uncertainty envolved.

\begin{figure}[th]
\setlength{\unitlength}{0.6mm}
\begin{center}
\begin{picture}(100, 80)
\put(-6, -3){\epsfxsize=7cm \epsfbox{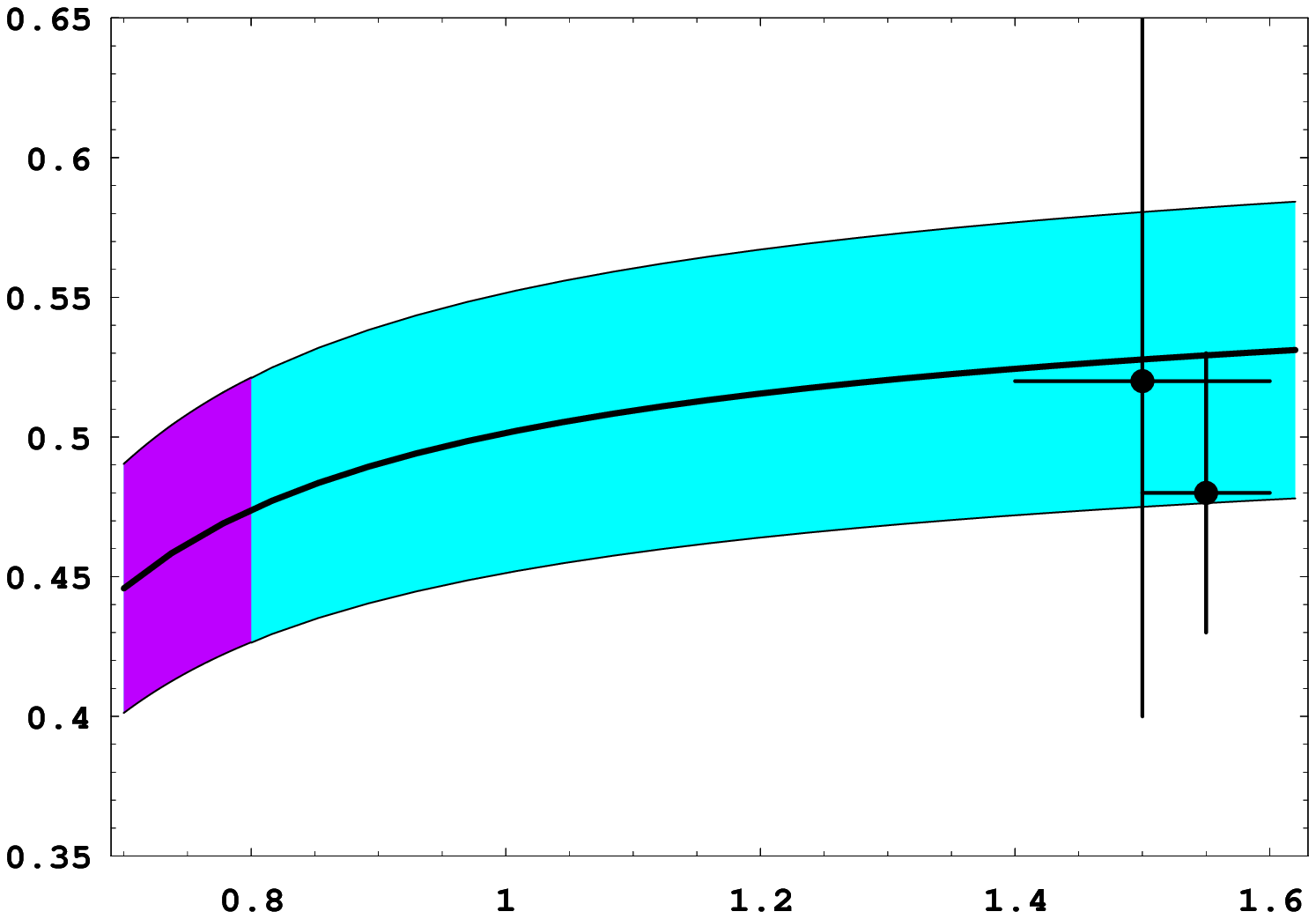}}
\put(85,-7){$\mu$, GeV}
\put(2, 80){$\tau, ps$}
\end{picture}
\end{center}
\caption{The dependence of $B_c$ meson lifetime on the scale $\mu$ in the
effective Hamiltonian (\ref{Heff}). The shad\-ed region shows the uncertainty
of estimates, the dark shaded region is the preferable choice as given by the
lifetimes of charmed mesons. The dots represent the values in the OPE
approach.} \label{Tau-fig}
\end{figure}

The width of beauty decay in the sum rules was derived using the similar
methods in \cite{Onish}: $\Gamma(B^+_c \rightarrow \bar c
c+X)=(28\pm5)\cdot10^{-14}~$GeV. The width of the electroweak annihilation is
taken from \cite{Bcstatus} as $12\cdot10^{-14}~$GeV. 

In Fig. \ref{Tau-fig} we present the $B_c~$meson lifetime calculated in the QCD
SR under consideration. We also show the results of the lifetime evaluation in
the framework of Operator Product Expansion in NRQCD \cite{Buchalla,BcOnish}. 

In contrast to OPE, where the basic uncertainty is given by the variation of
heavy quark masses, these parameters are fixed by the two-point sum rules for
bottomonia and charmonia, so that the accuracy of SR calculations for the total
width of $B_c$ is determined by the choice of scale $\mu$ for the hadronic weak
lagrangian in decays of charmed quark. We show this dependence in Fig.
\ref{Tau-fig}, where $\frac{m_c}{2} < \mu < m_c$ and the dark shaded region
corresponds to the scales preferred by data on the charmed meson lifetimes. The
discussion on the optimal choice of scale in hadronic decays is addressed in
\cite{full}.
We suppose that the preferable choice of scale in the $c\to s$ decays of
$B_c$ is equal to
$
\mu^2_{B_c} = \mu_{c\bar b} \cdot\mu_{c\bar s}\approx (0.85\; {\rm GeV})^2,
$
and at $a_1(\mu_{B_c}) =1.20$ in the charmed quark decays we predict
$
\tau[B_c] = 0.48\pm 0.05\;{\rm ps.}
$

\section{Conclusion}
We have investigated the semileptonic decays of $B_c$ meson due to the weak
decays of charmed quark in the framework of three-point sum rules in QCD. We
have pointed out the important role played by the Coulomb-like
${\alpha_s}/{v}$-corrections. As in the case of two-point sum rules, the
form factors are about three times enhanced due to the Coulomb renormalization
of quark-meson vertex for the heavy quarkonium $B_c$. We have studied the
dependence of form factors on the threshold energy, which determines the
continuum region of $\bar b s$ system. The obtained dependence has the
stability region, serving as the test of convergency for the sum rule method.
The HQET two-point sum rules for the leptonic constant $f_{B_s}$ and
$f_{B_s^*}$ have been reanalyzed to introduce the term caused by the product
of quark and gluon condensates. This contribution essentially improves the
stability of SR results for the leptonic constants of B mesons, yielding:
$f_B=140\div 170$ MeV.

We have studied the soft limit for the form factors in combined HQET/NRQCD
technique at the recoil momentum close to zero, which allows us to derive the
generalized relations due to the spin symmetry of effective lagrangian. The
relations  are in a good agreement with the full QCD results, which means that
the corrections to the form factors in both relative velocity of heavy quarks
inside the $\bar b c$ quarkonium and the inverse heavy quark masses are small
within the accuracy of the method. 
 
Next, we have studied the nonleptonic decays, using the assumption on the
factorization of the weak transition. The results on the widths and branching
fractions for various decay modes of $B_c$ are collected in Tables.

Finally, we have estimated the $B_c$ meson lifetime, and showed the dependence
on the scale for the hadronic weak lagrangian in decays of charmed quark
$
\tau[B_c] = 0.48\pm 0.05\;{\rm ps.}
$
Our estimates are in a good agreement with the theoretical predictions for the
lifetime in both the potential models and OPE as well as with the experimental
data.   

This work was in part supported by grants of RFBR 99-02-16558 and
00-15-96645.

\end{document}